\documentclass[12pt]{article}
\usepackage[all]{xy}
\usepackage{wrapfig}
\usepackage{epic}
\usepackage{amsmath}[1996/11/01]
\usepackage{amssymb,amsthm,amsfonts,latexsym,epsfig,graphics}
 \def\beql#1#2\eeql{\begin{equation}\label{#1}#2\end{equation}}
	 \usepackage{tikz}
	 \usetikzlibrary{positioning}
	 \usetikzlibrary{arrows}
	 \usetikzlibrary{decorations.pathreplacing}
	 \usetikzlibrary{decorations.markings}

\advance \textheight by -1 \topmargin
\advance \textheight by -1 \headheight
\advance \textheight by -1 \headsep
\advance \textheight by -2 \footskip
\vsize = \textheight
\hsize = \textwidth
\DeclareMathOperator{\id}{id}

\DeclareMathOperator{\charac}{char}

\DeclareMathOperator{\Ker}{Ker}

\newtheorem{theorem}{Theorem}[section]

\newtheorem{kor}[theorem]{Corollary}

\newtheorem{example}[theorem]{Example}

\newcommand{\bew}{\noindent\underline{Proof.}\ }

\newtheorem{remark}[theorem]{Remark}
\newtheorem{lemma}[theorem]{Lemma}

\newtheorem{definition}[theorem]{Definition}

\newcommand{\Z}{{\mathbb{Z}}}

\newcommand{\F}{{\mathbb{F}}}
\newcommand{\N}{{\mathbb{N}}}

\newcommand{\C}{{\mathcal{C}}}
\newcommand{\FG}{\F G}

\newcommand{\eb}{\phantom{zzz}\hfill{$\square $}\smallskip}

\title{Relative projective group codes over chain rings}
\author{Simon Eisenbarth \thanks{Lehrstuhl D f\"ur Mathematik, RWTH Aachen University,
52056 Aachen, Germany. Email: \texttt{simon.eisenbarth@rwth-aachen.de}}
\and Sihuang Hu\thanks{School of Cyber Science and Technology,
Shandong University, Qingdao, China. Email: \texttt{husihuang@sdu.edu.cn}}
\date{}
}
\begin{document}
\maketitle
\bibliographystyle{plain}


\medskip

\begin{abstract}
A structure theorem of the group codes which are relative projective for the subgroup $\lbrace 1 \rbrace$ of $G$ is given. With this, we show that all such relative projective group codes in a fixed group algebra $RG$ are in bijection to the chains of projective group codes of length $\ell$ in the group algebra $\FG$, where $\F$ is the residue field of $R$. We use a given chain to construct the dual code in $RG$ and also derive the minimum Hamming weight as well as a lower bound of the minimum euclidean weight.
\end{abstract}

\section{Introduction}

The theory of error-correcting codes was introduced by Golay and Hamming in 1949 and 1950, respectively. A natural goal of coding theory is to find interesting codes, i.e. codes with a high minimum distance, with a rich algebraic structure or other properties. One way to find such interesting codes are group codes. Let $\F$ be a finite field and $G$ be a finite group, then a group code is a left ideal in the group algebra $\FG$ (in this paper we only consider left ideals). They have a long history in coding theory, starting with S. Berman and  F.J. MacWilliams, who studied group codes over cyclic, abelian and dihedral groups (\cite{Berman69}, \cite{MacWilliams69}, \cite{MacWilliams70}).

Additionally, linear codes over rings have been studied increasingly in the last years. The research first focused on integer residue rings, especially $\Z / 4 \Z$, as some nonlinear binary codes are the image of linear codes over this ring under the so called Gray map (\cite{Ham94}), but also other rings were considered. Chain rings have received special attention. Their properties lie closest to the properties of finite fields, so it is expected that the structure of codes over these rings resembles those of classical coding theory. Furthermore, the class of finite chain rings contains some important infinite families of rings, for example the integer residue rings of prime power order, Galois rings, and certain group rings. In \cite{EN18} we used chain rings to classify self-dual codes with an automorphism whose order is the characteristic of the underlying field.

Recently, a result about linear complementary pairs (LCP) of group codes over fields has been extended to finite chain rings (\cite{LCP20}), but besides cyclic codes, group codes over (chain) rings have not been researched extensively.

In this paper, we characterize a certain type of group codes over finite chain rings. The idea is related to the construction in \cite{NebeHu}, where chains of cyclic codes are used to construct strongly perfect lattices. In the same way, we use a chain of projective group codes (i.e. the codes are generated by idempotents) to construct a group code over a chain ring. Projective group codes have been studied by many authors (for example \cite{FerrazMilies2005}, \cite{Nilpotent}, \cite{ComplementaryDual}), since all group codes are projective if the group algebra is semisimple and in some cases the generating idempotent can be used to determine the minimum distance and other properties. Our construction yields the group codes over a commutative Artinian chain ring which are relative projective for the subgroup $\lbrace 1 \rbrace$ of $G$, so they are in bijection to the chains of projective group codes with a certain length over the residue field. Such a chain can be used to derive a lower bound of the euclidian minimum distance of the resulting code.

\section{Chains of group codes and group codes over chain rings}

Throughout the paper let $R$ be a commutative Artinian chain ring with $1$, denote by $\mathfrak{m} = \pi R$ its maximal ideal and $\F := R / \mathfrak{m}$ the residue field, a finite field of characteristic, say $p$.  Let $\ell $ denote the length of $R$, i.e. the minimal
natural number such that $\mathfrak{m}^{\ell } = \lbrace 0 \rbrace$. The choice of the generator $\pi$ of the maximal ideal $\mathfrak{m}$ defines $R$-module isomorphisms
\begin{displaymath}
\alpha _j : \mathfrak{m}^j / \mathfrak{m}^{j+1}  \to \F , \pi^j r + \mathfrak{m}^{j+1} \mapsto r+ \mathfrak{m}
\end{displaymath}
for $j = 0, \dots, \ell-1$. We also fix a finite group $G$ and denote by $RG$ and $\FG$ the group rings of $G$ over $R$ respectively $\F$. So $RG$ is the $R$-algebra on the  free $R$-module with basis $G$ with the $R$-bilinear multiplication defined by the group multiplication. We extend the isomorphisms $\alpha _j$ to obtain $R$-module isomorphisms $\alpha _j : \mathfrak{m}^j RG / \mathfrak{m}^{j+1} RG  \to \FG$ $(j=0, \dots, \ell-1)$. Then $\alpha _0$ is an $R$-algebra homomorphism.

\begin{definition}
\begin{itemize}
\item[a)] A group code $C$ over $\F$ is a left ideal in $\FG$. We call $C$ a projective group code, if there exists an idempotent $e \in \FG$ such that $C = \FG e$.
\item[b)] A group code $\C$ over $R$ is a left ideal in $RG$. We call $\C$ a relative projective group code, if it is relative projective for the subgroup $\lbrace 1 \rbrace$ of $G$ in the sense of homological algebra, i.e. for any short exact sequence
\begin{displaymath}
0 \to \mathcal{M} \to \mathcal{N} \stackrel{\varphi}{\to} {\C} \to 0
\end{displaymath}
of $RG$-modules for which there is an $R$-module homomorphism $\psi : \C \to \mathcal{N}$ with $\varphi \circ \psi = \id_{\C}$, there is an $RG$-module homomorphism $\psi' : \C \to \mathcal{N}$ with $\varphi \circ \psi' = \id_{\C}$. In other words, if the sequence is right-split as $R$-module, it is already right-split as $RG$-module.
\end{itemize}
\end{definition}

Relative projective over a field is projective, but over a ring this is wrong, e.g. $R = \Z / 4 \Z$ and $\mathcal{C} = \Z / 2 \Z$. Note that if $p$ does not divide $|G|$, then all group codes in $\FG$ are projective and all group codes in $RG$ are relative projective.

\begin{remark}[{\cite[Proposition 2.1.6]{Zim14}}] \label{RemarkProjectiveSplit}
Since $RG$ is a free $R$-modul, a group code $\C$ over $R$ is relative projective if and only if the short exact sequence
\begin{displaymath}
0 \to \Ker(m) \to RG \otimes_{R} \C \stackrel{m}{\to} \C \to 0
\end{displaymath}
splits, where $m : \lambda \otimes c \mapsto \lambda c$ is the multiplication map.
\end{remark}

\begin{remark}
There is a well-known bijection between the idempotents in $\FG$ and the idempotents in $RG$ (see for example \cite[Theorem 7.3.5]{Web16}). Clearly, if $\epsilon \in RG$ satisfies $\epsilon^2 = \epsilon$ then also $e := \alpha_0(\epsilon) \in \FG$ is an idempotent. On the other hand given $e^2 = e \in FG$, we may chose a preimage $\epsilon_0 \in RG$ with $\alpha_0(\epsilon_0) = e$. Then $\epsilon_0^2 - \epsilon_0 \in \mathfrak{m} RG$ and putting
\begin{displaymath}
\epsilon_i := 3 \epsilon_{i-1}^2 - 2 \epsilon_{i-1}^3
\end{displaymath}
for $i = 1,\dots,\ell$ constructs the idempotent $\epsilon^2 = \epsilon := \epsilon_{\ell} \in RG$ with $\alpha_0(\epsilon) = e$.
\end{remark}

In the next theorems, we show that all relative projective group codes (independent on whether $p$ divides $|G|$ or not) can be constructed via nested chains of projective group codes over $\F$.

\begin{theorem}
Let
\begin{displaymath}
\C_{\star} : C_0 \leq C_1 \leq \dots \leq C_{\ell - 1}
\end{displaymath}
be a nested chain of projective group codes over $\F$ with generating idempotents $e_0, \dots, e_{\ell-1}$. Let $\epsilon_j \in RG$ be an idempotent with $\alpha_0(\epsilon_j) = e_j$ for all $0 \leq j \leq \ell -1$, then
\begin{displaymath}
\C := \hat{\C}_{\star} := RG \left( \sum_{j=0}^{\ell-1} \pi^j \epsilon_j \right)
\end{displaymath}
is a relative projective group code over $R$.
\end{theorem}
\bew
By remark \ref{RemarkProjectiveSplit}, $\C$ is relative projective if and only if the short exact sequence
\begin{displaymath}
0 \to \Ker(m) \to RG \otimes_{R} \C \stackrel{m}{\to} \C \to 0
\end{displaymath}
splits. Define the $R$-module homomorphism
\begin{displaymath}
h : \mathcal{C} \to RG \otimes_R \mathcal{C}, \sum_{g \in G} r_g g \cdot \sum_{j=0}^{l-1} \pi^j \epsilon_j \mapsto \sum_{g \in G} r_g \cdot g \otimes \sum_{j=0}^{l-1} \pi^j \epsilon_j.
\end{displaymath}
For an element $\sum_{g \in G} r_g g \cdot \sum_{j=0}^{l-1} \pi^j \epsilon_j \in \mathcal{C}$ we obtain
\begin{align*}
& m \left( h \left( \sum_{g \in G} r_g g \cdot \sum_{j=0}^{l-1} \pi^j \epsilon_j \right) \right) =  m \left( \sum_{g \in G} r_g \cdot g \otimes \sum_{j=0}^{l-1} \pi^j \epsilon_j \right) \\
= & \sum_{g \in G} r_g \cdot m \left( g \otimes \sum_{j=0}^{l-1} \pi^j \epsilon_j \right) = \sum_{g \in G} r_g g \cdot \sum_{j=0}^{l-1} \pi^j \epsilon_j
\end{align*}
hence $m \circ h = \id _\mathcal{C}$ and the sequence splits.
\eb

\begin{theorem} \label{TheoremDecomp}
Let $\C \leq RG$ be a relative projective group code. Then there exist primitive, orthogonal idempotents $\epsilon_i \in RG$ and $a_i \in \N_0$, such that
\begin{displaymath}
\C = \bigoplus_{i=1}^s \pi^{a_i} RG \epsilon_i.
\end{displaymath}
\end{theorem}
\bew
We prove the assertion with an induction by $\min \lbrace a \in \N \mid \mathfrak{m}^a \mathcal{C} = \lbrace 0 \rbrace \rbrace$. If $a = 1$, then $\mathcal{C} \leq \mathfrak{m}^l RG$ is a relative projective $RG / \pi RG$-module, hence projective over $\F$ and the assertion follows. Now let $a > 1$. In this case, replace $R$ by $R / \mathfrak{m}^a R$ and $RG$ by $\mathfrak{m}^{\ell-a} RG$, then w.l.o.g. we can assume $a = \ell$ and $C \not \leq \mathfrak{m} RG$. Given a decomposition $1 = \epsilon_1 + \dots + \epsilon_k$ of $1 \in RG$ in orthogonal, primitive idempotents $\epsilon_i \in RG$, we find
\begin{displaymath}
\mathcal{C} = \mathcal{C} \cdot 1 \subseteq \oplus_{i=1}^k RG \epsilon_i.
\end{displaymath}
There exists an $i$, such that $\mathcal{C} \epsilon_i \not \subseteq \mathfrak{m} RG \epsilon_i$, otherwise $\mathcal{C}$ would be contained in $\mathfrak{m} RG$. Now we want to show that for such an $i$ the equality $\C \epsilon_i = RG \epsilon_i$ holds. Since $C$ is relative projective, $\mathcal{C} \epsilon_i$ is as well because the short exact sequence
\begin{displaymath}
0 \to \Ker(m) \to \underbrace{RG \otimes_R \mathcal{C} \epsilon_i}_{\cong \oplus_{j=1}^k RG \epsilon_j \otimes_R \mathcal{C} \epsilon_i} \stackrel{m}{\to} \mathcal{C} \epsilon_i \to 0
\end{displaymath}
splits (remark \ref{RemarkProjectiveSplit}). The element $e_i := \epsilon_i + \mathfrak{m} RG$ is an idempotent in $\FG$ whose lift is $\epsilon_i$. Then $\alpha_0( \C ) e_i$ is a projective module in $\FG e_i$ and because  $e_i$ must be primitive the equality $\alpha_0( \C ) e_i = \FG e_i$ holds, therefore $\C \epsilon_i = RG \epsilon_i$. Now consider the short exact sequence
\begin{displaymath}
0 \to \Ker(\varphi) \to \C \stackrel{\varphi}{\to} \C \epsilon_i = RG \epsilon_i  \to 0,
\end{displaymath}
where $\varphi$ is the multiplication with $\epsilon_i$. Since $RG \epsilon_i$ is a free $R$-module and relative projective, the sequence splits as $R$-module so it already splits as $RG$-modules. By the splitting lemma $\C$ is isomorphic to $\Ker(\varphi) \oplus RG \epsilon_i$ and the argument can be applied to $\Ker(\varphi)$.
\eb

\begin{lemma}
Let $\C \leq RG$ be a relative projective group code. For all $0 \leq j \leq \ell -1$ define
\begin{displaymath}
C_j := \alpha_j \left( \frac{\C \cap \mathfrak{m}^j RG}{\C \cap \mathfrak{m}^{j+1} RG} \right) \leq \FG.
\end{displaymath}
Then
\begin{displaymath}
\C_{\star} : C_0 \leq C_1 \leq \dots \leq C_{\ell -1}
\end{displaymath}
is a chain of projective group codes over $\F$.
\end{lemma}
\bew
By theorem \ref{TheoremDecomp} there exist primitive, orthogonal idempotents $\epsilon_i \in RG$ and $a_i \in \N_0$, such that
\begin{displaymath}
\C = \bigoplus_{i=1}^s \pi^{a_i} RG \epsilon_i.
\end{displaymath}
In particular we have $C \cap \mathfrak{m}^j RG = RG f_j$ with $f_j = \sum_{a_i \leq j} \pi^j \epsilon_i + \sum_{a_i > j} \pi^{a_i} \epsilon_i$. We obtain
\begin{displaymath}
\alpha_j \left( \frac{\C \cap \mathfrak{m}^j RG}{C \cap \mathfrak{m}^{j+1} RG}  \right) = \FG \underbrace{ \left( \sum_{a_i \leq j} \alpha_o( \epsilon_i ) \right) }_{ =: e_j }.
\end{displaymath}
As a sum of orthogonal, primitive idempotents, all $e_j$ are idempotents in $\FG$ which satisfy $e_i e_j = e_{\min(i,j)}$, so $C_i \leq C_j$ for $i \leq j$.
\eb

\begin{kor}
Relative projective group codes over $R$ are in bijection to chains of length $\ell$ of projective group codes over $\F$.
\end{kor}

\subsection{Duality}

The bilinear form
\begin{displaymath}
( \cdot , \cdot ) : \FG \times \FG \to \F , \left( \sum_{g \in G} a_g g , \sum_{g \in G} b_g g \right) \mapsto \sum_{g \in G} a_g b_g
\end{displaymath}
on $\FG$ (and completely analogous for $RG$) is symmetric, non-degenerate and $G$-invariant. The dual code of some group code $C \leq \FG$ is defined as
\begin{displaymath}
C^{\perp} = \lbrace v \in FG \mid (c,v) = 0 \text{ for all } c \in C \rbrace.
\end{displaymath}

\begin{remark}[{\cite[Corollary 1.3]{Wil02}},{ \cite[Remark 4.2]{Wil02}}] \label{LemmaSelfDual}
Let ${}^* : \FG \to \FG, g \mapsto g^{-1}$ be the conjugation map of $\FG$ and let $C = \FG e$ be a projective group code. Then the dual code is given by $C^{\perp} = \FG (1-e^*)$. A self-dual group code can not be projective as the complement is not an ideal. The group algebra $\FG$ contains a self-dual group code if and only if $\charac(\F) = 2$ and the order of $G$ is even.
\end{remark}

\begin{lemma}
Let $\C \leq RG$ be a relative projective group code over $R$ with
\begin{displaymath}
\C_{\star} : C_0 \leq C_1 \leq \dots \leq C_{\ell-1}.
\end{displaymath}
Then the corresponding chain for the dual code is given by $\C^{\perp}_{\star} = C_{\ell-1}^{\perp} \leq \dots \leq C_1^{\perp} \leq C_0^{\perp}$.
\end{lemma}
\bew
Let $b_1, \dots, b_m \in \C \cap \mathfrak{m}^j RG$, such that $(b_i + \C \cap \mathfrak{m}^{j+1} RG )_{i=1,\dots,m}$ is an $R$-generating set of $\frac{\C \cap \mathfrak{m}^j RG}{\C \cap \mathfrak{m}^{j+1}RG}$. Then we have
\begin{displaymath}
\alpha_{\ell-1-j} \left( \frac{\langle b_1,\dots,b_m \rangle^{\perp} \cap \mathfrak{m}^{\ell-1-j} RG}{\langle b_1,\dots,b_m \rangle^{\perp} \cap \mathfrak{m}^{\ell-j} RG} \right) = C_j^{\perp} \leq \FG
\end{displaymath}
and the assertion follows.
\eb

\begin{remark}
If $\ell$ is even, a trivial example of a self-dual relative projective group code over $R$ is given by
\begin{displaymath}
\C_{\star} = \underbrace{ \lbrace 0 \rbrace \leq \dots \leq \lbrace 0 \rbrace }_{\ell / 2} \leq \underbrace{ \FG \leq \dots \leq \FG }_{\ell /2 }.
\end{displaymath}
If $\ell$ is odd, such a code must satisfy
\begin{displaymath}
\C_{\star} = C_0 \leq \dots \leq C_{\frac{\ell-1}{2}} = C_{\frac{\ell-1}{2}}^{\perp} \leq C_{\frac{\ell-3}{2}}^{\perp} \leq \dots \leq C_0^{\perp},
\end{displaymath}
but by remark \ref{LemmaSelfDual} the self-dual code $C_{\frac{\ell-1}{2}}$ can not be generated by an idempotent. Thus self-dual relative projective group codes over $R$ exist if and only if $\ell$ is even.
\end{remark}

\subsection{Minimum Distance}

For a group code $\mathcal{C} \leq RG$ the Hamming weight of a code word $c  = \sum_{g \in G} c_g g \in \mathcal{C}$ is defined as
\begin{displaymath}
w_H(c) := | \lbrace g \in G \mid c_g \neq 0 \rbrace |
\end{displaymath}
and the Hamming minimum distance is then
\begin{displaymath}
d_H(\mathcal{C}) = \min \lbrace w_H(c) \mid 0 \neq c \in \mathcal{C} \rbrace.
\end{displaymath}

\begin{theorem}
Let $\mathcal{C} \leq RG$ be a relative projective group code with $C_{\star} : C_0 \leq \dots \leq C_{\ell -1}$.
Then we have $d_H(\mathcal{C}) = d_H(C_{\ell-1})$.
\end{theorem}
\bew
The inequality $d_H(\mathcal{C}) \leq d_H( C_{\ell -1})$ is clear. Assume there exists a code word $0 \neq c \in \C$, such that $w_H(c) < d_H( C_{\ell -1})$. Let $i$ be minimal with $\pi^i c \in \pi^{\ell -1} RG$. In particular it is $\pi^i c \neq 0$ and $w_H(\pi^i c ) \leq w_H(c)$. On the other hand, we have $\alpha_{\ell -1}( \pi^i c ) \in C_{\ell -1}$ and $w_H( \alpha_{\ell -1}(\pi^i c)) = w_H(\pi^i c)$, so $d_H(C_{\ell -1}) \leq w_H(c) < d_H( C_{\ell -1})$, a contradiction.
\eb

In the special case $R = \Z / p^{\ell} \Z$ the euclidian weight of code word $c = \sum_{g \in G} c_g g \in (\Z / p^{\ell} \Z) G$ is defined as
\begin{displaymath}
w_E(c) := \min \left\lbrace \sum_{g \in G} a_g^2 \mid a_g \in \Z, a_g + p^{\ell} \Z = c_g \right\rbrace
\end{displaymath}
and the euclidian minimum weight is therefore
\begin{displaymath}
d_E(\C) = \min \lbrace w_E(c) \mid 0 \neq c \in \C \rbrace.
\end{displaymath}

\begin{theorem}[see {\cite[Theorem 2.8]{NebeHu}} for a similar proof]
Let $R = \Z / p^{\ell} \Z$ and let $\C \leq RG$ be a relative projective group code with $\C_{\star} : C_0 \leq \dots \leq C_{\ell-1}$. If there exists an $\gamma > 0$ with $d_E(C_j) \geq \frac{\gamma}{p^{2j}}$ for every $j = 0,\dots,\ell-1$, then $d_E(\C) \geq \gamma$.
\end{theorem}
\bew
Let $c = \sum_{g \in G} (c_g + p^{\ell} \Z) g \in \C$ and let $j$ be maximal such that $c \in \pi^j RG$. Then
\begin{displaymath}
\sum_{g \in G} (y_g + p \Z)g \in C_j \text{ with } y_g = c_g / p^j \in \Z.
\end{displaymath}
As $\sum_{g \in G} y_g^2 \geq d_E(C_j) \geq \frac{\gamma}{p^{2j}}$ we have $\sum_{g \in G} c_g = p^{2j} \sum_{g \in G} y_g^2 \geq p^{2j} \frac{\gamma}{p^{2j}} = \gamma$, hence $d_E(\C) \geq \gamma$.
\eb

\begin{example}
We consider the ring $R = \Z /  4 \Z$ and the Dihedral group $G = D_{2n}$. The self-dual relative projective group codes over $R$ are given by the chains
\begin{displaymath}
\C_{\star} : C_0 \leq C_0^{\perp}
\end{displaymath}
of self-orthogonal, projective group codes over $\F_2$. The idempotents in $\F_2 G$ can be easily calculated if the structure of the group ring $\F_2 G$ is known, which is closely related to the factorization of $x^n - 1$ in $\F_2[x]$.
In \cite{DHS01} the maximal possible minimum Hamming weight of a self-dual code over $\Z / 4 \Z$ is given (up to length $24$), the following table compares those values with the maximal minimum Hamming distance of self-dual, relative projective group codes in $RG$ (second row). Espacially for odd $n$ the Dihedral codes are often optimal.

\begin{figure}[!h]
\centering
\begin{tabular}{c|cccccccccccccccc}
$2n$ & $10$ & $12$ & $14$ & $16$ & $18$ & $20$ & $22$ & $24$ & $26$ & $28$ & $30$ & $32$ & $36$ & $38$ & $40$ & $42$ \\
\hline
       & $2$  & $2$  & $3$  & $3$  & $4$  & $4$  & $6$  & $8$  &      &      &      &      &      &      &      &      \\
       & $2$  & $2$  & $3$  & $1$  & $4$  & $4$  & $6$  & $3$  & $5$  & $5$  & $6$  & $1$  & $6$  & $7$  & $5$  & $8$
\end{tabular}
\end{figure}
\end{example}

\end{document}